\documentclass[11pt,a4paper, superscriptaddress, nofootinbib]{revtex4-1}
\pdfoutput=1
\usepackage{amsmath}
\usepackage{graphicx}
\usepackage{caption}
\usepackage{placeins}
\usepackage{float}
\usepackage[english]{babel}
\usepackage{epstopdf}

\begin{document}

\title{Stabilising the Planck mass shortly after inflation}
\author{Carsten van de Bruck}
\email[Email address: ]{C.vandeBruck@sheffield.ac.uk}
\affiliation{Consortium for Fundamental Physics, School of Mathematics and Statistics, University of Sheffield, Hounsfield Road, Sheffield, S3 7RH, United Kingdom}

\author{Adam Christopherson}
\email[Email address: ]{achristopherson@ufl.edu}
\affiliation{Department of Physics, University of Florida, Gainesville, FL 32611, USA}

\author{Mathew Robinson}
\email[Email address: ]{app11mrr@sheffield.ac.uk}
\affiliation{Consortium for Fundamental Physics, School of Mathematics and Statistics, University of Sheffield, Hounsfield Road, Sheffield, S3 7RH, United Kingdom}

\begin{abstract}
We consider a model of the early universe which consists of two scalar fields: the inflaton, and a second field which drives the stabilisation of the Planck mass (or gravitational constant). We show that the non-minimal coupling of this second field to the Ricci scalar sources a non-adiabatic pressure perturbation. By performing a fully numerical calculation we find, in turn, that this boosts the amplitude of the primordial power spectrum after inflation.
\end{abstract}
\keywords{Inflationary Cosmology, Early Universe Cosmology}

\maketitle%

\section{Introduction}

Cosmological observations put strong constraints on processes which could have occurred in the early universe. For example, models of inflation are tested with the properties of the cosmic microwave background radiation (CMB), such as the CMB anisotropies, CMB polarisation, non--Gaussianity and spectral distortions to the black--body spectrum. The {\sc Planck} satellite \cite{Ade:2013uln, Planck:2015xua} provides the most recent observational data of the CMB. Non--Gaussian statistics originating from inflationary physics can furthermore be probed with studies of the large scale structures (LSS) in the universe \cite{Desjacques:2010jw}.
This is one of the goals of state-of-the-art current and future experiments, such as the Dark Energy Survey \cite{Abbott:2005bi}, the Large Synoptic Survey Telescope (LSST) \cite{Ivezic:2008fe} and the {\sc Euclid} satellite \cite{Amendola:2012ys}.

The number of different inflationary models is vast. The simplest models consist of a single scalar field minimally coupled to gravity. However, the phenomenology of even these simple models is rich, with hundreds of different choices for the inflationary potential \cite{Martin:2014vha}. More complex models arise from including more than one scalar field, which can lead to qualitative differences. These differences occur due to fluctuations not just in one field direction, but now in more than one direction (i.e. isocurvature or non-adiabatic pressure perturbations, to which we will return later). Then, one could consider a single field with a non-standard kinetic term, such as k-inflation \cite{Garriga:1999vw}; these such models are often motivated by theories with extra dimensions, e.g., DBI inflation from string theory \cite{Alishahiha:2004eh}. For more complexity, the scalar field could have a non-minimal coupling to gravity, such as in the newest version of the Higgs inflation model \cite{Bezrukov:2007ep} (however, for the majority of cases, this can be treated as a field with a minimal coupling by moving to the Einstein frame and modifying the potential). Finally, the most complex inflationary models contain multiple scalar fields non-minimally coupled to gravity, and with non-standard kinetic terms \cite{Kaiser10a, Kaiser:2013sna}. 

By comparing each model's predictions with observational data, we can rule out regions of model space, with the ultimate goal to obtain a single inflationary model which best fits the data. Recent data provides bounds on the gravitational wave signature for which the simplest single field inflationary model with an $m^2\phi^2$ potential is disfavoured \cite{Ade:2015tva}. Therefore, it is particularly important to continue to investigate the dynamics and observational predictions of inflationary models beyond the simplest single scalar field model. One interesting model not belonging to the single field class is the curvaton model \cite{Enqvist:2001zp,Lyth:2001nq,Moroi:2001ct}. This model consists of a second field, the curvaton, in addition to the inflaton. The curvaton is dynamically unimportant during inflation, but its fluctuations source the curvature perturbation.

In this paper, we address the question of whether a possible stabilisation of the Planck mass (or gravitational constant) just after inflation can have a sizeable effect on the primordial power spectrum of the curvature perturbation. In theories in which the four--dimensional Planck mass are not constant, its dynamics is usually driven by one or several moduli fields. These describe for example the size of the extra--dimensional space. Since the time evolution of the gravitational constant is strongly constrained by experiments (see e.g. \cite{Will:2014xja} for a recent update on experimental tests of General Relativity), any stabilisation of the moduli field(s) must have happened in the early universe\footnote{Alternatively, the post-inflation evolution of the field(s) could be very slow.}. The stabilisation could have happened well before inflation ended, affecting scales well outside the visible horizon. If the stabilisation happened during the last 60 e--folds of inflation, possible signatures in the curvature perturbation power--spectra can be produced \cite{Ashoorioon:2014yua}. If the stabilisation happened later, in the radiation dominated epoch, the rapid oscillations of the scalar field(s) can affect the expansion history \cite{Steinhardt:1994vs,Perivolaropoulos:2002pn}. In the scenario discussed in this paper, Newton's constant stabilised a few e--folds after inflation. We take into account the possibility that the moduli field driving the evolution of Newton's constant can decay into other particles as well. Our setup is therefore related to the standard curvaton scenario. In the absence of a non-adiabatic pressure perturbation, the curvature perturbation $\zeta$ on uniform density hypersurfaces is known to be conserved on superhorizon scales \cite{Wands2}. However, this is not necessarily the case if several fields are dynamically important. Even if inflation has ended, the decay of fields at a later stage can significantly influence the evolution of $\zeta$ (see e.g. \cite{Enqvist:2001zp,Lyth:2001nq,Moroi:2001ct,Ashcroft:2004rs,Bassett:1998wg, Bassett:1999mt, Bassett:1999ta}). In the case of a scalar field driving the evolution of the Planck mass, we find that the non--minimal coupling to the Ricci scalar can boost the amplitude of the curvature perturbation by several orders of magnitude, even if the Planck mass varies only by a very small amount. 

The paper is structured as follows: in the next section, we present the model, and the governing equations for the background and perturbations of the model. Then, in Section~\ref{sec:numerics} we describe our numerical procedure, before presenting results in Section~\ref{sec:results}. Finally, we conclude in Section~\ref{sec:conclusion}.

\section{Theory and analytical results}
\label{sec:theory}

The model we consider consists of two scalar fields, namely of the inflaton $\phi$ and the field $\sigma$, which describes the evolution of the Planck mass. In the Jordan frame, the action is given by
\begin{eqnarray}
\label{eq:action}
\mathcal{S} = \int \mathrm{d}^4 x \sqrt{-g} \left[ \frac{1}{2}f(\sigma)R - \frac{1}{2}g^{\mu\nu}\left(\partial_{\mu}\phi\partial_{\nu}\phi + \partial_{\mu}\sigma\partial_{\nu}\sigma\right)  - V(\phi, \sigma) + \mathcal{L}_{\rm int} \right]\,,
\end{eqnarray}
where $g_{\mu\nu}$ is the metric tensor, $V(\phi, \sigma)$ the potential and $\mathcal{L}_{\rm int}$ is the interaction Lagrangian, describing the perturbative decay \cite{Huston13} of both the $\phi$ and $\sigma$ into radiation. By working in the Jordan frame, the decay rates can be calculated in the standard way. We denote them by $\Gamma^\phi$ and $\Gamma^\sigma$ respectively. Since we are interested in the effect of stabilising the Planck mass, we expand $f(\sigma)$ around its minimum, keeping only the leading term:  
\begin{eqnarray}
f(\sigma) = 1 + \frac{\alpha}{2}(\sigma - \sigma_{\rm min})^2\,.
\end{eqnarray}
We denote the masses of the fields by $m_\phi$ and $m_\sigma$ and assume that the fields are not directly interacting. Note that we are working in units with reduced Planck mass $M_{\rm PL}=1$.Therefore, the potential is given by 
\begin{equation}
V(\phi,\sigma) = \frac{1}{2}m_\phi^2 \phi^2 + \frac{1}{2}m_\sigma^2 \sigma^2. 
\end{equation}

To consider the evolution of cosmological perturbations, we work in the longitudinal gauge, in which the line element takes the form \cite{Bardeen80,Kodama84,Mukhanov:1990me}
\begin{eqnarray}
ds^2 = -(1+2\Phi)dt^2 + a^2(t)\left(1-2\Psi\right)\delta_{ij}dx^i dx^j~.
\end{eqnarray}
Here, $a(t)$ is the scale factor, $\Phi$ and $\Psi$ are independent metric perturbations, which depend on all coordinates. In the Jordan frame, $\Phi$ and $\Psi$ are not equal even in the absence of anisotropic stress (see Eq. (\ref{PhiPsi}) below). 
The equations of motion for the system can be obtained by varying the action in Eq.~(\ref{eq:action}). In the background, we have evolution equations for the two scalar fields
\begin{eqnarray}
\ddot{\phi} &=& - V_\phi - 3H\dot{\phi} -\Gamma^\phi\dot{\phi}\,,\\
\ddot{\sigma} &=& - V_\sigma  - 3H\dot{\sigma} + R f_\sigma/2 - \Gamma^\sigma\dot{\sigma} \label{curvaton} \,, 
\end{eqnarray}
in addition to an energy conservation equation for the radiation fluid
\begin{equation}
\dot{\rho_\gamma} = -4H\rho_\gamma + \Gamma^\phi\dot{\phi}^2+  \Gamma^\sigma\dot{\sigma}^2\,,
\end{equation}
and the Friedmann equation
\begin{equation}
H^2 = \frac{1}{3f}\left[ \frac{\dot{\phi}^2}{2} + \frac{\dot{\sigma}^2}{2} + V + \rho_\gamma \right] - \frac{f_\sigma \dot{\sigma}H}{f}
\end{equation}
In these equations, a subscript $\phi$ or $\sigma$ denotes a partial derivative with respect to the field, and we have written the derivatives with respect to cosmic time, $t$. We shall also use the slow roll parameter defined by \cite{Liddle:2000cg},
\begin{equation}
\epsilon\equiv-\frac{\dot{H}}{H^2}\,,
\end{equation}
in order to more simply write the Ricci scalar, which can be expressed as $R = 6H^2(2-\epsilon)$ along with its perturbation
\begin{equation}
\delta R = -6\ddot{\Psi} - 6H(\dot{\Phi} + 4\dot{\Psi}) - 2R\Phi + 2\frac{k^2}{a^2}(\Phi - 2\Psi)\,.
\end{equation}

Considering now the linear perturbations, we obtain a Klein-Gordon equation for each field \cite{Kaiser10a}
\begin{align}
\ddot{\delta\phi} &= -3H\dot{\delta\phi} - \left( \frac{k^2}{a^2} + V_{\phi\phi}\right)\delta\phi - V_{\phi\sigma}\delta\sigma +  \dot{\phi}(\dot{\Phi} + 3\dot{\Psi}) - 2V_\phi\Phi \,,\\
\label{dsigdashdash}
\ddot{\delta\sigma} &= -3H\dot{\delta\sigma} - \left( \frac{k^2}{a^2} + V_{\sigma\sigma} - \frac{f_{\sigma\sigma}R}{2}\right)\delta\sigma - V_{\sigma\phi}\delta\phi +  \dot{\sigma}(\dot{\Phi} + 3\dot{\Psi}) - 2V_\sigma\Phi + \frac{f_\sigma}{2}(2R\Phi + \delta R) \,,
\end{align}
along with a conservation equation for the radiation fluid
\begin{equation}
\dot{\delta\rho_\gamma} = -4H\delta\rho_\gamma + 4\rho_\gamma\dot{\Phi}- 2\frac{k^2}{a^2}(\dot{\Psi} + H\Phi) + 2\Gamma_\phi(\dot{\phi}\dot{\delta\phi} - \frac{\dot{\phi}^2}{2}\Phi) + 2\Gamma_\sigma(\dot{\sigma}\dot{\delta\sigma} - \frac{\dot{\sigma}^2}{2}\Phi)\,.
\end{equation}
The metric perturbation $\Psi$ satisfies the following evolution equation
\begin{align}
\ddot{\Psi} &= - 3H\dot{\Psi} - H\dot{\Phi} - H^2(3-2\epsilon)\Phi  \nonumber\\
& + \frac{1}{2f} \Bigg[\dot{\phi}\dot{\delta\phi} + \dot{\sigma}\dot{\delta\sigma} - (\dot{\phi}^2 + \dot{\sigma}^2)\Phi - V_\phi\delta\phi - V_\sigma\delta\sigma - 2\ddot{f}\Phi - \dot{f}(\dot{\Psi} + 2H\Phi) \Bigg. \nonumber\\
&\left.  \qquad \quad- \frac{\delta f}{f}\left( \frac{\dot{\phi}^2}{2} + \frac{\dot{\sigma}^2}{2}  - V  + \ddot{f} + 2H\dot{f}  \right) + \ddot{\delta f} + 2H\dot{\delta f} + \frac{k^2}{a^2}\delta f \right] 
 \label{dPsidashdash}
\end{align}
along with the constraint
\begin{equation}\label{PhiPsi}
\Phi = \Psi - \frac{\delta f}{f}\,.
\end{equation}

The predictions from inflationary models are mapped onto observations (such as the temperature fluctuations of the CMB) in a simple way by introducing a curvature perturbation. The curvature perturbation on uniform density hypersurfaces, $\zeta$, is defined as
\begin{equation}
\zeta=-\Psi-\frac{H}{\dot{\rho}}\delta\rho\,,
\end{equation}
where here, $\rho$ and $\delta\rho$ are the energy density and perturbation for the entire matter content of the universe. We can obtain an evolution equation for $\zeta$ which, in the large-scale limit, takes the form
\begin{equation}
\dot{\zeta}=-\frac{H}{(\rho+P)}\delta P_{\rm nad}\,,
\end{equation}
where the non-adiabatic pressure perturbation, $\delta P_{\rm nad}$, is defined as
\begin{equation}
\delta P_{\rm nad}\equiv\delta P-\frac{\dot{P}}{\dot{\rho}}\delta\rho\,.
\end{equation}
For a minimally coupled single field model of inflation (or, equivalently, for a universe containing a single fluid), the curvature perturbation, $\zeta$, is conserved for both canonical and non-canonical models of inflation, independent of the theory of gravity \cite{Wands2,Rigopoulos:2003ak,Christopherson:2008ry}. This allows us to compare inflationary predictions directly to observations by mapping the field fluctuations onto the curvature perturbation. Since it is conserved, we do not need to worry about the mechanism by which inflation ends and the universe reheats.  

 However, moving beyond these simple models, the non-adiabatic pressure (or entropy) perturbation is non-zero, and therefore the curvature perturbation can continue to evolve and be enhanced on super-horizon scales. This feature has been exploited in numerous scenarios containing multiple minimally coupled scalar fields (see, e.g.,
 Refs.~\cite{GarciaBellido:1995qq,Bassett:2005xm, Bassett:1998wg, Bassett:1999mt, Bassett:1999ta,Huston:2011fr, Huston13}). In these models, we must take into account the reheating phase in order to make reliable predictions. Models with non-minimally coupled scalar fields, on the other hand, will produce a distinct source of entropy perturbations, arising due to the coupling of the scalars. In the model we present above, it is expected that, after inflation has ended and during the reheating phase, these entropy perturbations can become sizeable due to the fact that $\dot{f}$ and $\ddot{f}$ no longer need to remain small \cite{Kaiser10a}. It is these non-adiabatic pressure perturbations, and the resulting amplification of the power spectrum, that we will investigate in the remainder of the paper. 
 
\begin{table}
\caption{\label{tab:notation}A table clarifying our notation for the subscripts denoting various stages in the evolution of $\sigma$.}
\begin{tabular}{|l||l|}
\hline
$\sigma_{\rm{ini}}$  & The initial value of $\sigma$ \\
 \hline
$\sigma_{\rm{end}}$ & The value $\sigma$ reaches at the end of inflation, before rolling down to its minimum  \\ \hline
$\sigma_{\rm{min}}$    & The minimum in the expansion of $f(\sigma)$ \\ \hline
\end{tabular}
\end{table}
 
We will solve the system of equations derived above. It will be necessary to  follow the evolution of the secondary field, $\sigma$, throughout the inflationary phase, the decay of the inflationary field, $\phi$, and the radiation epoch right up until $\sigma$ itself has decayed and no longer contributes to the overall energy density of the universe. This is important so as not to restrict ourselves to the case where 
the auxiliary field starts at its minimum, $\sigma_{\rm min}$ (see Table \ref{tab:notation} for subscript notation), and is pushed away, but to also include cases where the field evolves towards $\sigma_{\rm min}$ during inflation. It is often the case that when studying subdominant, curvaton-like fields, the calculation begins in a post-inflationary radiation-dominated phase and proceeds from there. For the usual, minimally coupled fields this is sufficient, since 
 $\sigma$ does not evolve until late on, after the end of inflation \cite{Lyth:2001nq}. However, as this no longer holds in our case, we must track it throughout. It is still important for this coupling to remain small so as not allow $\sigma$ to contribute too much and impact upon the dynamics of inflation itself. 
 
This feature, which distinguishes our setup from the standard curvaton scenario,
arises due to the explicit coupling to the Ricci scalar causes the field to obtain an effective mass, which might not be small compared to $H^2$. In the slow--roll approximation we find that $\sigma$ evolves according to 
\begin{equation}
\sigma \propto e^{3\alpha\frac{2-\epsilon}{3-\epsilon} N},
\end{equation} 
where $\epsilon$ is assumed to be roughly constant and $N$ is the e--fold number. This equation follows directly from Eq. ({\ref{curvaton}}), writing this equation in terms of e--fold number $N$ and neglecting the bare mass of the field. Therefore, it is often the case that the field--value of $\sigma$ at horizon crossing is different from the value of $\sigma$ at the end of inflation.

\section{Numerical method}
\label{sec:numerics}
We will solve the governing equations derived in the previous section using a code written in 
Python. This starts at the beginning of inflation and runs right through to the end of the decay of the second field, when the power spectrum has reached its final value. This is the full numerical simulation including field perturbations, their gravitational counterparts and those in the radiation fluid created in the final stages. We begin by running through the background in order to ascertain the values needed to set up the initial conditions for each mode, $k$, such that $k_* = 50a_{\rm i} H_{\rm i}$. We then set the initial conditions of the perturbations as those of the Bunch-Davies vacuum \cite{Bunch:1978yq} at this point and begin the full perturbation code. The perturbation equations are then each integrated twice, independently, by first setting the initial value of $\delta\sigma$ to zero whilst leaving $\delta\phi$ to take its Bunch-Davies vacuum form:

\begin{eqnarray}
\delta\phi, \delta\sigma &=& \frac{e^{-ik\tau}}{\sqrt{2ka_{\rm i}}}\,,\\
\delta\phi', \delta\sigma' &=& \frac{-ik e^{-ik\tau}}{\sqrt{2ka_{\rm i}^3 H_{\rm i}^2}}\,,
\end{eqnarray}
and then vice versa.  We normalize the number of e-foldings to be $N = 0$ at horizon exit, and so plot our results around this value.

The code is split into four sections, each solved successively with the end values to each one used as the initial conditions in the next:
\begin{enumerate}
  \item {\bf Inflation}: \textnormal{This covers the period from $N=0$ through to when the inflaton crosses the minimum of the potential, at which point we switch the decay, $\Gamma_\phi$ on.}
 \item {\bf Inflaton decay}: \textnormal{Covering the period through the first part of reheating, but before the secondary field has begun to decay.}
 \item {\bf Overlap decay}: \textnormal{The inflaton still contributes a significant amount to the overall energy density but the secondary field, $\sigma$ too has started to decay, so we switch on $\Gamma_\sigma$.}
\item {\bf Secondary field decay}: \textnormal{Finally, we switch off the evolution of the $\phi$ field altogether as it is so difficult and time consuming to follow the vastly different scales involved in both this and much smaller secondary field, $\sigma$. This then continues until the power spectra settles on a specific value and all the energy density of the universe is held within the radiation.}
\end{enumerate}
We take the decay parameters to be $\Gamma^\phi=10^{-8}M_{\rm PL}$ and $\Gamma^\sigma=10^{-14}M_{\rm PL}$, and the masses of  the scalar fields to be
$m_\phi=10^{-7}M_{\rm PL}$ and $m_\sigma=10^{-10}M_{\rm PL}$. These values are chosen to be close to those in Ref.~\cite{Huston13}, and compatible with the limits in Eq. (11) of Ref.~\cite{Bartolo:2002vf}

While the first two sections can be integrated in a matter of minutes, the latter sections can take some considerable time to track the oscillatory phases throughout decay of the secondary field. This is due to the small scales involved in comparison to the inflationary phases. Since the sudden decay approximation does not hold in this case \cite{Lyth:2002my}, the later sections are crucial in our numerical procedure in order to obtain an accurate result.
Even in the standard case, with $\alpha = 0$, we find that the ratio of $\sigma$ to the other components in the universe, $r_{\rm{dec}}$, evaluated when $H = \Gamma^\sigma$ is not the true point at which $r_{\rm{dec}}$ reaches its maximum value. An improved (and increased) value can be attained slightly before this at $H \sim 3 - 5\Gamma^\sigma$ or, more accurately still, read off from its numerical maximum. The need for following the decay in full becomes even more apparent when we look at the results, in the next section. We find that oscillations in the non-adiabatic pressure perturbation towards the end of curvaton decay play an important role too.

\section{Results}
\label{sec:results}

\subsection{The case: $\sigma_{\rm{min}} = \sigma_{\rm{ini}}$}
\label{sec:results1}

\begin{figure}
\includegraphics[width=\linewidth]{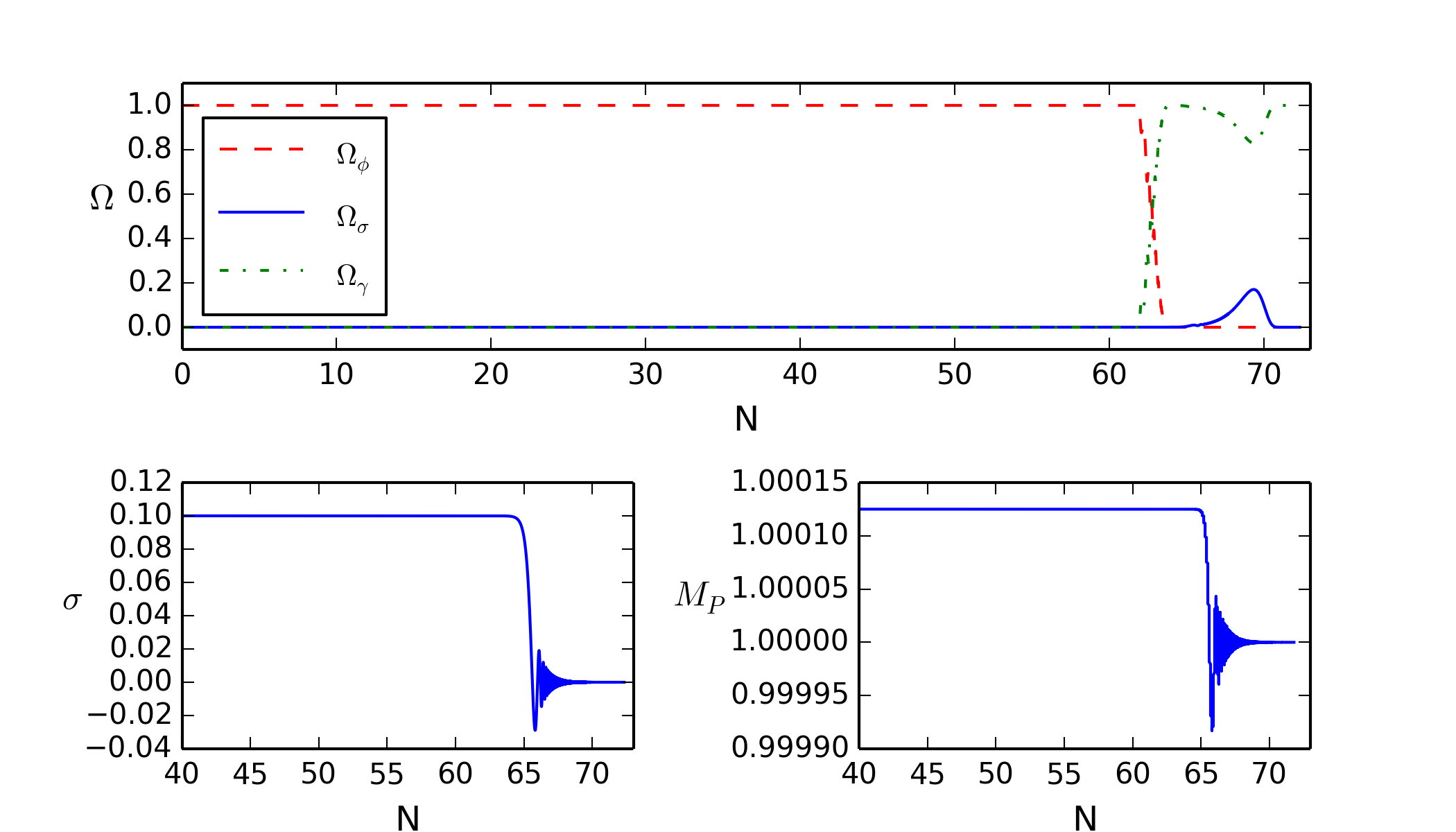}
  \caption{{\it Top}: The evolution of the relative energy density in each species, $\Omega_i$ for both the $\alpha = 0$ and $\alpha = -0.005$ cases, which overlap throughout. {\it Bottom left}: The background evolution of $\sigma$ for both $\alpha$ values and for $\sigma_{\rm{min}} = \sigma_{\rm{ini}} = 0.1$. {\it Bottom right}: The evolution of the effective Planck mass for $\alpha = -0.005$ when $\sigma_{\rm{min}} = \sigma_{\rm{ini}} = 0.1$.}
  \label{om_-0.005}
\end{figure}

\begin{figure}
\includegraphics[width=\linewidth]{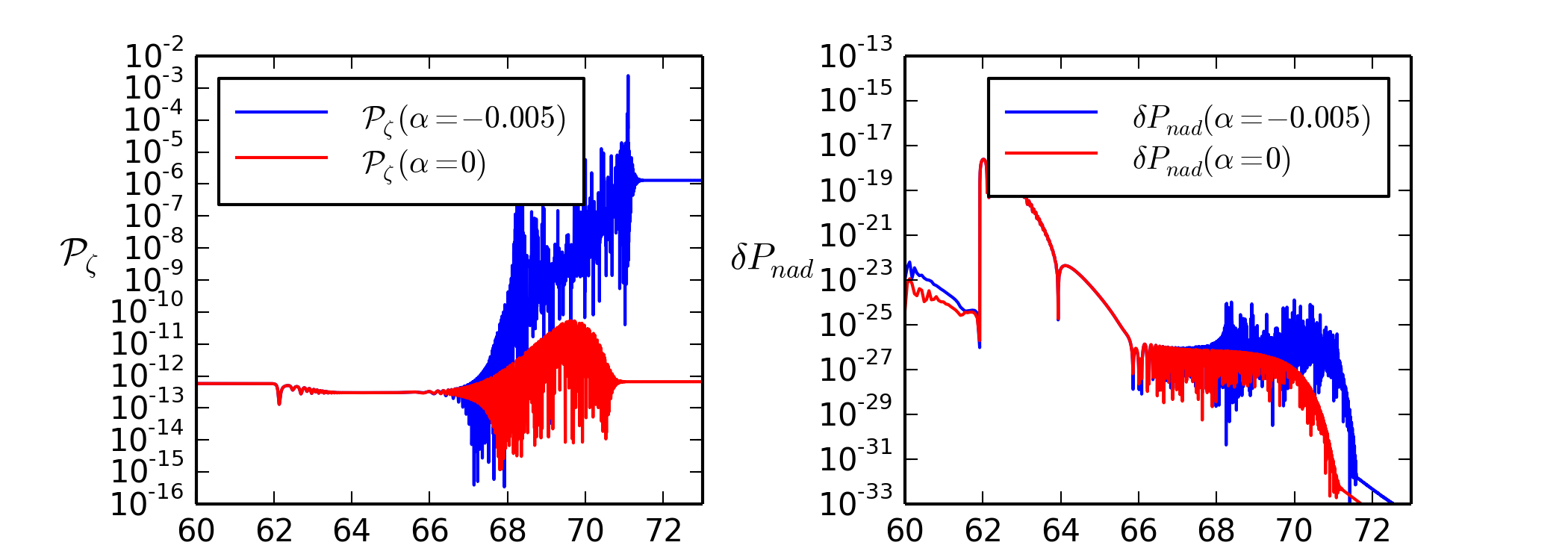}
  \caption{The power spectrum of the curvature perturbation ({\it left}) for both $\alpha = 0$ and $\alpha = -0.005$ cases and the associated $\delta P_{\rm nad}$ ({\it right}).}
  \label{Pdnad1}
\end{figure}

For this section we set $\sigma_{\rm{min}} = \sigma_{\rm{ini}}$ in order to exclude any evolution of the secondary field during inflation (see Figure~\ref{om_-0.005}). By keeping $\sigma_{\rm{min}} = \sigma_{\rm{ini}}= 0.1$ and comparing to a standard curvaton scenario, for which $\alpha = 0$ and $r_{\rm{dec}} \simeq 0.18$ (Figure~\ref{om_-0.005}), we see a significant change in the amplitude of the final power spectrum for a given $k$. In Figure~\ref{Pdnad1} we plot the final twelve efolds as the inflaton decays, followed by the radiation dominated and secondary field dominated phases in terms of both $P_{\mathcal{\zeta}}$ (the dimensionless power spectrum) and $\delta P_{\rm{nad}}$. In this and later plots in the paper, we take $k=0.05 \,{\rm Mpc}^{-1}$.  This shows the influence of the non-adiabatic pressure on the final power spectra; the $\delta P_{\rm{nad}}$ survives for around an efold longer and has a maximum amplitude of up to roughly 100 times that of the standard case. Figure~\ref{om_-0.005} shows that this increase in amplitude is not due to a more dominant secondary field, as the value of $\Omega_\sigma$ at the time of decay remains roughly constant (the change is of the order of $0.1\%$). 

\begin{figure}
\includegraphics[width=\linewidth]{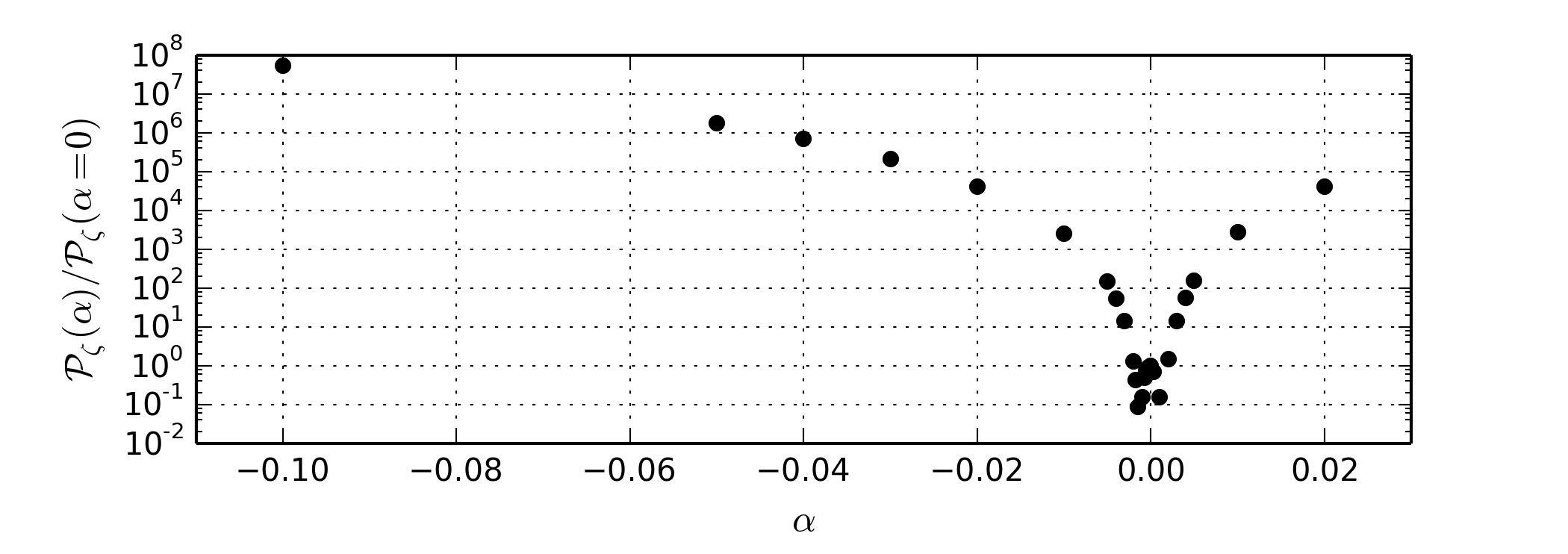}
  \caption{The amplitude of the power spectrum as a function of $\alpha$ normalised to the $\alpha=0$ power spectrum: ${\mathcal{P}}_{\mathcal{\zeta}}(\alpha)/{\mathcal{P}}_{\mathcal{\zeta}}(0)$.}
  \label{Pvasym}
\end{figure}

For the case with $\sigma_{\rm{ini}} = 0.1$, as above, we see in Figure~\ref{Pvasym} that the dependence on $\alpha$ is independent of sign. This might be expected due to the boost in power spectrum coming as the secondary field oscillates and decays. Terms such as $\dot{f}$ and $\ddot{f}$, which both contribute to $\delta P_{\rm{nad}}$, effectively average out as their sign changes back and forth. For this reason, in the case of $\sigma_{\rm{min}} = \sigma_{\rm{ini}}$ we shall now only look at the effect of $\lvert \alpha\rvert$. For the cases when $\sigma_{\rm{min}} \neq \sigma_{\rm{ini}}$, which we will consider in the next section, this may no longer remain true as the sign of $\alpha$ will introduce a scaling of the final field value, $\sigma_{\rm{end}}$, which can in turn affect the final power spectrum. We also observe a slight dip on either side of $\alpha = 0$, for which the amplitude decreases before increasing again. We have checked that this is not a numerical artefact. We do not have a physical explanation for this dip and the complexity of the governing equations makes it difficult to address this question analytically. 

\begin{figure}
\includegraphics[width=\linewidth]{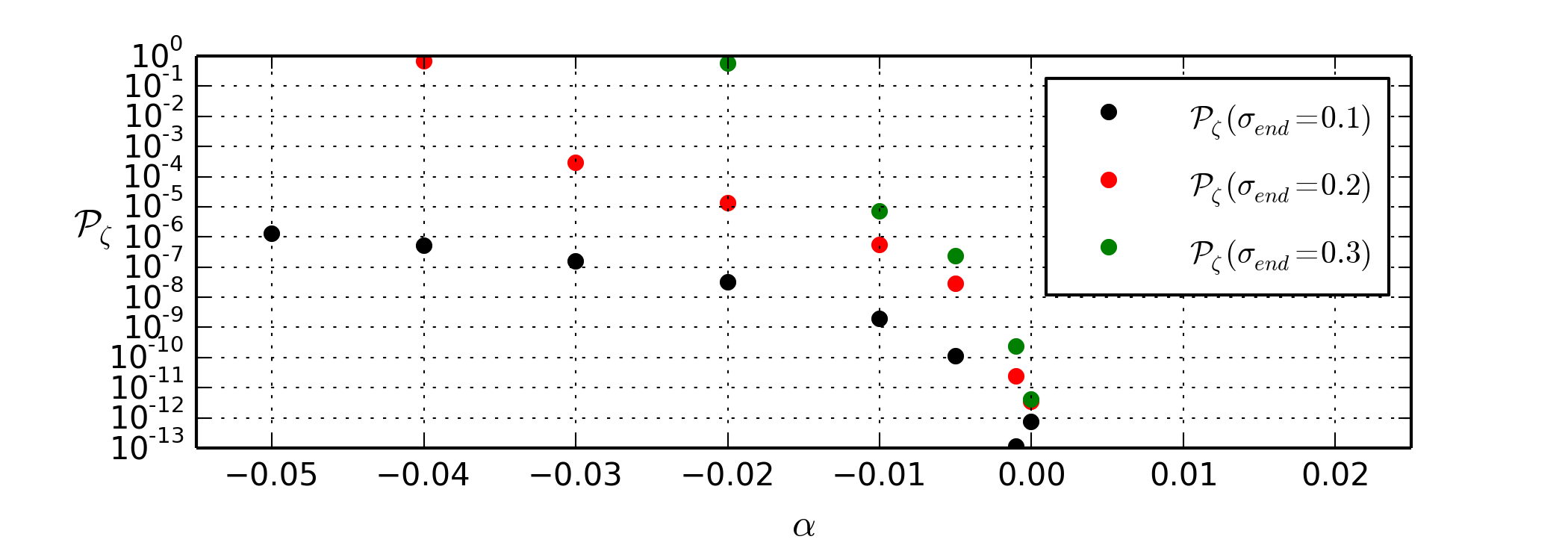}
  \caption{Amplitude of the power spectrum ${\mathcal{P}}_{\mathcal{\zeta}}$ as a function of $\alpha$ for three different values of $\sigma_{\rm end} = 0.1,0.2,0.3$}
  \label{alphas1}
\end{figure}

Finally, in these simple cases, it is useful to compare how $\sigma_{\rm {end}}$ affects the outcome for various values of $\lvert \alpha\rvert$. We will specifically focus on $\sigma_{\rm{end}} = 0.1,0.2,0.3$ which results in $r_{\rm{dec}}$ values of $0.17, 0.45$ and $0.62$ respectively.  For these final values of $\sigma$ we would expect a varying increase in the amplitude of the power spectrum arising simply from the standard curvaton results. At the end of inflation we find $P_{\mathcal{\zeta}}= 3.01\times 10^{-13}$ and this value is boosted by factors of $2.51, 11.13$ and $13.85$ respectively by the end of curvaton decay with $\alpha = 0$. In Figure~\ref{alphas1} this small boost is apparent in the values at $\alpha = 0$ but is insignificant in comparison to the subsequent amplitude increases as we increase $\alpha$ from $0$. The results for each $\sigma_{\rm{end}}$ diverge for increasing $\alpha$ due to the fact that for each $\sigma_{\rm{end}}$ we also have $\sigma_{\rm{min}} = \sigma_{\rm{end}}$, so that the difference between the true minimum ($ \sigma \simeq 0$) and the local minimum associated with $f(\sigma_{\rm min})$  increases as the values for $\sigma_{\rm end}$ increase.

\subsection{The case: $\sigma_{\rm min} \neq \sigma_{\rm ini}$}

In the more general case we have two possibilities, namely $\alpha > 0$ and  $\alpha < 0$. This choice  plays a role in the evolution of $\sigma$ during inflation which can in turn affect $\sigma_{\rm end}$ and $r_{\rm dec}$. By choosing  $\alpha$ to be negative, we can pull $\sigma$ towards its local minimum before decay; a positive $\alpha$ has the opposite effect and pushes it away. This second case soon becomes unworkable for values of $\alpha$ approaching $0.05$ or greater due to the exponential increase apparent in the value of $\sigma$. Due to the symmetry of $\alpha$ shown in Figure~\ref{Pvasym} this need not be too concerning, however, as once $\sigma$ reaches its final value at the end of inflation we can still study the subsequent effects purely by using negative values. The only benefit of using $\alpha > 0$ comes in the ability to further vary the trajectory of $\sigma$ to test that our results are independent of it. This can be done by setting $\sigma$ to a number of different initial values and using $\alpha$ to control its final value in order to compare results. 

\begin{figure}
\includegraphics[width=\linewidth]{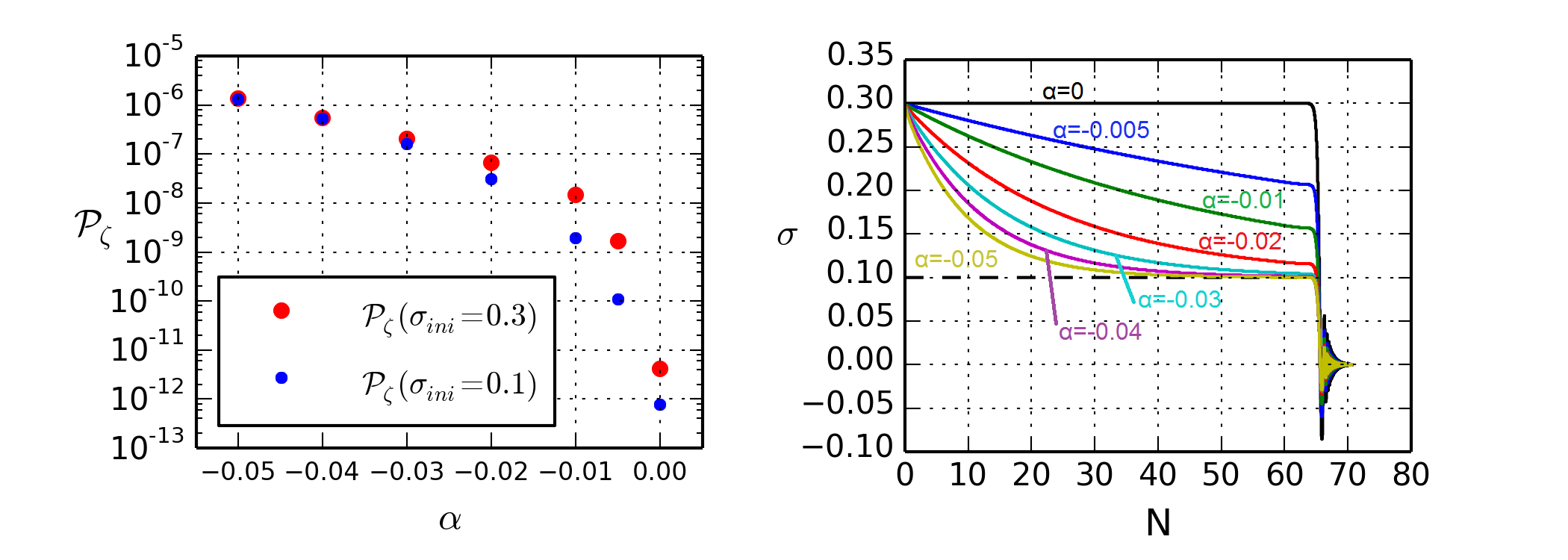}
  \caption{{\it Left:} The power spectra for varying $\alpha$ with $\sigma_{\rm ini} = 0.3$ and $\sigma_{\rm min} = 0.1$ (red) in comparison to the case of $\sigma_{\rm ini} = \sigma_{\rm min} = 0.1$ (blue). {\it Right}: The background trajectories for $\sigma$ for each of these cases. }
  \label{min01}
\end{figure}

In each case we find that the power spectrum is dominated by the value of $\sigma_{\rm min}$ with a lesser but still noticeable dependence on $\sigma_{\rm end}$. This is most simply demonstrated by Figure~\ref{min01} in which we show two cases, both with $\sigma_{\rm min} = 0.1$ but with $\sigma_{\rm ini} = 0.1$ and $0.3$ respectively. We let $\alpha$ run over the same values used previously which, for $\sigma_{\rm ini} = 0.3$, gives various values of $\sigma_{\rm end}$:  $0 > \alpha > -0.03$  results in $0.3 > \sigma_{\rm end} > 0.1$, while  $\alpha < -0.03$ gives $\sigma_{\rm end} = 0.1$ (see the right hand side of Figure~\ref{min01}). It is clear from the left hand side of Figure~\ref{min01} that while $\alpha$ remains small, the final amplitude of the power spectrum differs from that of the $\sigma_{\rm ini} = \sigma_{\rm min}$ case. This can be explained by the observation that in these cases the field has not had enough time to reach $\sigma_{\rm end} = 0.1$ due to the smallness of $\alpha$. For larger $\alpha$, however, the two cases converge because $\sigma_{\rm end}$ now equals $0.1$ in each of these examples. 

\subsection{The case: $\sigma_{\rm min} = 0$}
Finally, we give an example which demonstrates both that the evolution of $\sigma$ throughout inflation has little to no impact (other than the dependence on $\sigma_{\rm end}$) on the final power spectra amplitudes and that we get no noticeable boost when $\sigma_{\rm min} = 0$. Here we take $\sigma_{\rm ini} = 0.05$ and $\alpha = \{0.0056, 0.011, 0.0145\}$ which gives $\sigma_{\rm end} = \{0.1, 0.2, 0.3\}$, respectively. From these we find that the amplitude of the final power spectrum is multiplied by factors of $1.03, 1.29$ and $1.51$. However, in comparison to the factors involved in the standard curvaton-like case given in Section~\ref{sec:results1}, for these values of $\sigma_{\rm end}$ we see that the changes represent additional increases of only $1-3\%$. These are insignificant when taking into account the usual approximations inherent in the curvaton model and those increases found earlier in the paper for $\sigma_{\rm min} \neq 0$.

\section{Conclusion}
\label{sec:conclusion}

In this paper we have studied a model of the early universe consisting of two scalar fields: the inflaton and a second field which controls the stabilisation of the Planck mass. We work in the Jordan frame, for which the second field is non-minimally coupled to gravity; this choice allows us to deal with the decay of the fields in the usual way. We have investigated numerically the effects of this coupling on the power spectrum of primordial fluctuations. 
It has previously been shown that a non-minimally coupled scalar field can induce changes in the curvature perturbation on super-horizon scales via the introduction of terms proportional to $f, \dot{f}$ and $\ddot{f}$ in the non-adiabatic pressure perturbation, $\delta P_{\rm nad}$ \cite{Kaiser10a}. Here, we have quantified this effect. We have shown that it can play an important role on the amplitude of the power spectrum in a non-minimally coupled curvaton-like case, in which the secondary field decays only after inflation is complete. Allowing the effective Planck mass to evolve in such a way, even by the smallest of amounts, leads to dramatic changes in the amplitude of the final power spectrum in comparison to the standard curvaton scenario.

The effect of this amplitude boost can also be linked to the spectral index, $n_s$ and tensor-scalar ratio, $r_{TS}$ by parameterising the power spectrum as \cite{Enqvist:2013paa}
\begin{eqnarray}
\mathcal{P}_\zeta = \mathcal{P}_\zeta^{(\phi)}+ \mathcal{P}_\zeta^{(\sigma)} = (1+R)\mathcal{P}_\zeta^{(\phi)}\,,
\end{eqnarray}
where
\begin{eqnarray}
R = \frac{\mathcal{P}_\zeta^{(\sigma)}}{\mathcal{P}_\zeta^{(\phi)}}\,.
\end{eqnarray}
This gives
\begin{eqnarray}
n_s - 1 = -2\epsilon + 2\eta_\sigma - \frac{4\epsilon-2\eta_\phi}{1+R} \qquad\text{and}\qquad r_{TS} = \frac{16\epsilon}{1+R} \,,
\end{eqnarray}
using the usual definitions of the slow roll parameters, evaluated at horizon crossing. The important point to note here is that  $n_s$ and $r_{TS}$ depend only on the ratio, $R$, not the mechanism by which the curvaton, or curvaton-like field, sources the final curvature perturbation. This places tight constraints on the values that $\alpha, \sigma_{\rm min}$ and hence $f(\sigma)$ can take according to the latest {\sc Planck} data \cite{Byrnes:2014xua}. We soon find ourselves with a spectral index approaching 0.98 -- as in the pure curvaton limit -- as $R$ becomes large with relatively small changes in $f$. This is also largely independent of any evolution in $\sigma$ during inflation because the inflaton dominates the universe at horizon crossing, when both the slow roll parameters are evaluated and the tensor perturbations freeze in. 

It remains to be seen whether a similar scenario to the one discussed will arise from fundamental theories of particle physics. If so, it will have an impact on inflationary model building in such theories. 

\acknowledgments
CvdB is supported by the Lancaster-Manchester-Sheffield Consortium for Fundamental Physics under STFC grant ST/L000520/1. AJC is supported by the U.S. Department of Energy under Grant No. DE-FG02-97ER41029 and, during the early stages of this work, was supported by the Sir Norman Lockyer Fellowship of the Royal Astronomical Society.

\end{document}